\documentclass[aps,prl,twocolumn,groupedaddress,floatfix]{revtex4}
\usepackage{amssymb,amsmath}
\usepackage{graphicx}
\usepackage{subfigure}
\usepackage[english]{babel}
\usepackage{float}
\usepackage{color}

\newcommand{\be}{\begin{equation}}
\newcommand{\ee}{\end{equation}}

\begin{document}

\title{The magnetoinductive dimer }

\author{Mario I. Molina}

\affiliation{Departamento de F\'{\i}sica, MSI-Nucleus on Advanced Optics,  and Center for Optics and Photonics (CEFOP), 
Facultad de Ciencias, Universidad de Chile, Santiago, Chile.\\
Tel: 56-2-9787275, Fax: 56-2-2712973, email: mmolina@uchile.cl}

\begin{abstract}
We examine a nonlinear magnetoinductive dimer and compute its
linear and nonlinear symmetric, antisymmetric and asymmetric modes in closed-form, in the rotating-wave approximation. A linear stability analysis of these modes reveals that the asymmetric mode is always stable, for any allowed value of the coupling parameter and for both, hard and soft nonlinearity. 
A numerical computation of the dimer dynamics reveals 
a magnetic energy selftrapping whose threshold increases for increasing dimer coupling.
\end{abstract}

\maketitle

\clearpage

Metamaterials are novel artificial materials characterized for having 
unusual electromagnetic wave propagation  properties, such as 
a negative dielectric permittivity and negative magnetic permeability over 
a finite frequency range. This feature makes them attractive for use as a constituent in negative refraction index materials\cite{negative refraction}.
A subclass of those metamaterials, the magnetic metamaterials (MMs), exhibit significant magnetic properties and negative magnetic response up to terahertz and optical frequencies\cite{21,22}.

One of the most well-known MMs consist of a metallic composite structure consisting of arrays of split-ring resonators (SRRs). The theoretical treatment of such structures relies mainly on the effective-medium approximation where the composite is treated as a homogeneous and isotropic medium, characterized by effective macroscopic parameters. The approach is valid, as long as the wavelength of the electromagnetic field is much larger than the linear dimensions of the MM constituents \cite{SRR1,SRR2,SRR3}.

The shortest array is the dimer, and consists on two SRRs coupled inductively. In spite of its simplicity, a dimer is capable of rich phenomenology including magnetic energy transfer, mode localization and even chaos\cite{PLA_LMTK}. 
Linear magnetic dimers have been used as constituent 
units for envisioned three-dimensional metamaterials or `stereometamaterials'
\cite{Liu}. The dynamics of magnetic energy localization in an asymmetric nonlinear dimer with dissipation and driving has been examined in ref.\cite{PLA_LMTK}, where it was concluded that asymmetry  
gives rise to chaotic dynamics and is also necessary for strong localization in one of the dimer sites. Also, the linear magnetic dimer with gain/loss terms constitutes yet another finite system with PT-symmetry properties, possessing a parameter window inside which its dynamics is bounded\cite{mm_PLA}.

In this work we focus on a nonlinear dimer system, consisting of two identical SRRs, and examine the exchange and localization of magnetic energy via a closed-form computation of the linear and nonlinear modes. We also examine the dynamics of selftrapping of magnetic energy as a function of the mutual coupling between the rings.

\begin{figure}[t]
\noindent
\includegraphics[scale=.5,angle=0]{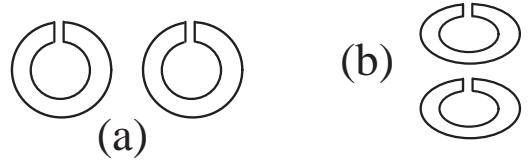}
\caption{Magnetoinductive dimer. In (a) $M<0$, while in (b) $M>0$.}
\label{fig1}
\end{figure}

Let us consider an array consisting of  two identical split ring resonators (SRRs), as shown in Fig.\ref{fig1}. The system is characterized by the values of the self-inductance $L$, linear capacitance $C_{l}$, mutual inductance $M$, characteristic voltage $U_{c}$ and linear dielectric constant $\epsilon_{l}$.
In the presence of dissipation and driving, the coupled equations for the time evolution of the charges $Q_{1}, Q_{2}$ are given approximately by
\begin{eqnarray}
L {d^{2} Q_{1}\over{ d t^{2}}} + M {d^{2} Q_{2}\over{ d t^{2}}} + R {d Q_{1}\over{d t}} + {Q_{1}\over{C_{l}}} & = &\nonumber\\
		- {\alpha\over{3 \epsilon_{l}}} {1\over{U_{c}^{2}}}- \left( {Q_{1}\over{C_{l}}} \right)^{2}  \epsilon_{0} \sin(\Omega t)&=0&\nonumber\\
L {d^{2} Q_{2}\over{ d t^{2}}} + M {d^{2} Q_{1}\over{ d t^{2}}} + R {d Q_{2}\over{d t}}+ {Q_{2}\over{C_{l}}} & &\nonumber\\
		- {\alpha\over{3 \epsilon_{l}}} {1\over{U_{c}^{2}}}- \left( {Q_{2}\over{C_{l}}} \right)^{2} \epsilon_{0} \sin(\Omega t)&=0&\label{eq:1}
\end{eqnarray}
The most commonly used dimer configurations are in-a-plane and on-an-axis (Fig.\ref{fig1}). In the first case, the mutual inductance is negative, while on the second case it is positive.

\begin{figure}[t]
\noindent
\includegraphics[scale=0.45,angle=0]{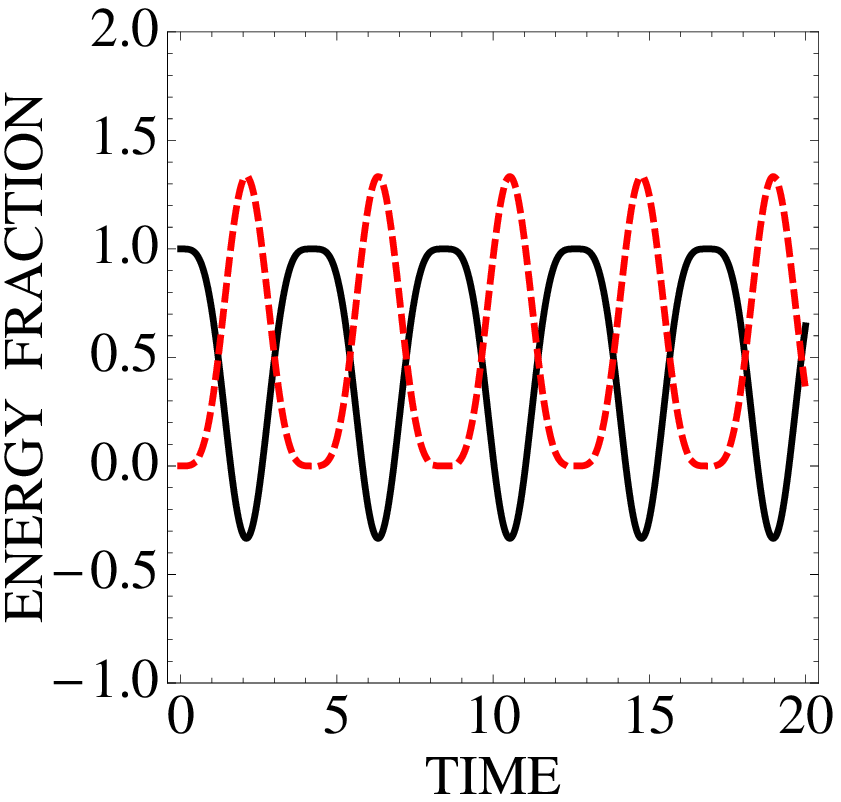}\hspace{0.4cm}
\includegraphics[scale=0.45,angle=0]{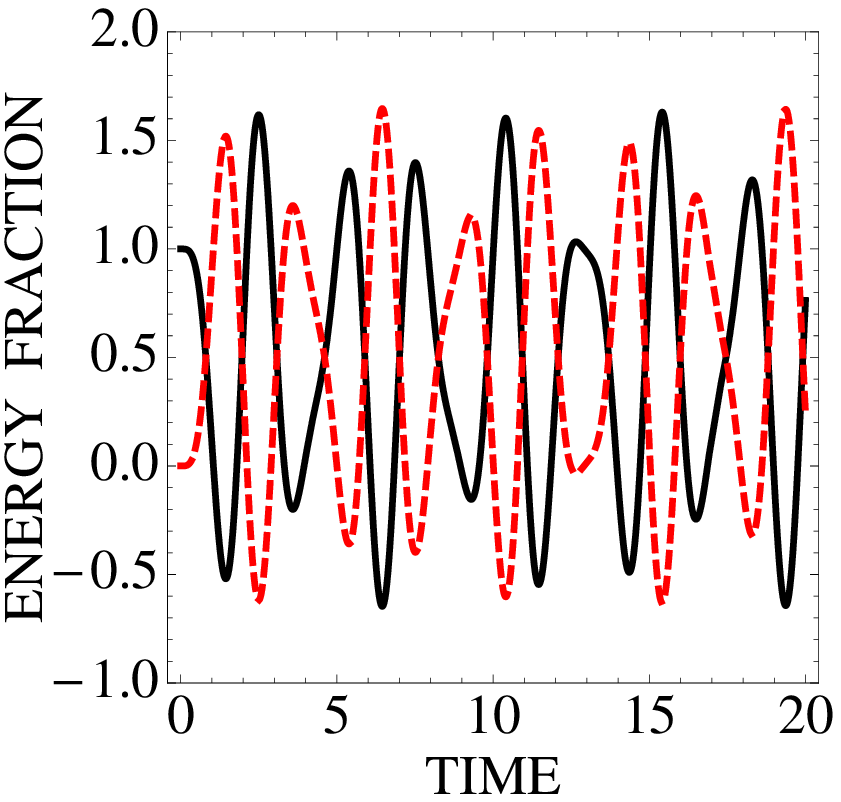}
\caption{Evolution of SRR energies for two coupling values: 8/10 (left) and 9/10 (right). The solid (dashed) curve denotes 
$H_{1} (H_{2})$. In the first case, the evolution is periodic; in the second it is quasiperiodic.}
\label{fig2}
\end{figure}

{\em Linear case}.\ We start by examining the simplest case, where there is no dissipation and driving force, and nonlinear effects are negligible. With the definitions
$Q_{c}=C_{l} U_{c}, \omega_{0}=1/\sqrt{L C_{l}},\tau=\omega_{0} t, \lambda=M/L<1$ and $q=Q/Q_{c}$, Eqs.(\ref{eq:1}) become:
\begin{eqnarray}
{d^{2}\over{ d \tau^{2}}} ( q_{1} + \lambda q_{2} ) + q_{1} & =& 0\nonumber\\
{d^{2}\over{ d \tau^{2}}} ( q_{2} + \lambda q_{1} ) + q_{2} & =& 0\label{eq:2}
\end{eqnarray}
It is easy to see that $H=(1/2)({\dot q_{1}}^2+{\dot q_{2}}^2)+(1/2)(q_{1}^2+q_{2}^2)+\lambda {\dot q_{1}}{\dot q_{2}}$ is a constant of motion, $d H/dt =0$. We define the energy contents $H_{1}$ and $H_{2}$ of each SRR as
\begin{equation}
H_{1}={1\over{2}}({\dot q_{1}}^2 + q_{1}^2 +
\lambda {\dot q_{1}}{\dot q_{2}})
\end{equation}
\begin{equation}
H_{2}={1\over{2}}({\dot q_{2}}^2 + q_{2}^2  + \lambda {\dot q_{1}}{\dot q_{2}}).
\end{equation}
We now look for stationary solutions: $q_{1,2}\sim \exp(i \beta \tau)$. After inserting into (\ref{eq:2}), one obtains $\beta=\pm 1/\sqrt{1+\lambda}, \pm 1/\sqrt{1-\lambda}$. As a direct application, we consider the initial-value problem $q_{1}(0)=q_{0}, {\dot q_{1}}(0)=0,q_{2}(0)=0,{\dot q_{2}}(0)=0$, whose solution is
\begin{eqnarray}
q_{1}(\tau)&=&{q_{0}\over{2}}\left[ \cos\left({\tau\over{\sqrt{1+\lambda}}}\right) + \cos\left({\tau\over{\sqrt{1-\lambda}}}\right) \right]\\\label{eq:q1}
q_{2}(\tau)&=&{q_{0}\over{2}}\left[ \cos\left({\tau\over{\sqrt{1+\lambda}}}\right) - \cos\left({\tau\over{\sqrt{1-\lambda}}}\right) \right]\label{eq:q2}
\end{eqnarray}
Since the two frequencies involved, $1/\sqrt{1\pm \lambda}$ are in general incommensurable, the motion of $q_{1}(\tau)$ and $q_{2}(\tau)$ will be quasiperiodic, unless $\lambda$ can be written as $(p^{2}-q^{2})/(p^{2}+q^{2})$, where $p,q$ are integers. Once in possession of $q_{1}(\tau)$ and $q_{2}(\tau)$ in closed form, we could write $H_{1}$ and $H_{2}$ explicitly as function of time, but the expressions are rather cumbersome and not particularly illuminating. Figure \ref{fig2} shows examples of the evolution of the SRRs energies for similar coupling values that lead to qualitatively different evolutions.
Besides the typical presence of quasiperiodicity, the SRRs energies are not positive-definite.
We see that even in this simple case, the dynamics is considerably more complex than for the linear case of the Discrete Nonlinear Schr\"{o}dinger (DNLS) dimer. In particular, it is not clear now how to define a proper coupling time. This is due to the fact that energy is not only stored as charges in the capacitors of the SRRs, but also in the magnetic fields across the SRRs slits.


{\em Nonlinear case}.\  In the presence of nonlinear effects, Eqs.(\ref{eq:2}) become
\begin{eqnarray}
{d^{2}\over{ d \tau^{2}}} ( q_{1} + \lambda q_{2} ) + q_{1} - \chi\ q_{1}^{3} & = & 0\nonumber\\
{d^{2}\over{ d \tau^{2}}} ( q_{2} + \lambda q_{1} ) + q_{2} - \chi\ q_{2}^{3}  & =& 0\label{eq:3}
\end{eqnarray}
where $\chi=\alpha/3 \epsilon_{l}$. In order to have an oscillatory behavior at all, we need that $q_{1,2}^{2}<1/\chi$ for $\chi>0$. For $\chi<0$ no such restriction is necessary, since the potential is always hard in that case.
We see that the change $\lambda\rightarrow -\lambda$ is equivalent to
$q_{1}\rightarrow q_{1}, q_{2}\rightarrow -q_{2}$ or $q_{1}\rightarrow -q_{1}, q_{2}\rightarrow q_{2}$. We could call this the ``staggered'' mode. More interestingly, Eq.(\ref{eq:3}) admits the conserved quantity:
\begin{equation}
H = {1\over{2}}({\dot q_{1}}^2 + {\dot q_{2}}^2 + q_{1}^2 + q_{2}^2) +
\lambda {\dot q_{1}}{\dot q_{2}} -{\chi\over{4}} (q_{1}^{4}+q_{2}^{4}),
\end{equation}
which we call the magnetic energy. The individual energy contents of each ring are then
\begin{eqnarray}
H_{1} &=& {1\over{2}}({\dot q_{1}}^2 + q_{1}^2 +
\lambda {\dot q_{1}}{\dot q_{2}}) -{\chi\over{4}} q_{1}^{4}\nonumber\\
H_{2} &=& {1\over{2}}({\dot q_{2}}^2 + q_{2}^2 +
\lambda {\dot q_{2}}{\dot q_{1}}) -{\chi\over{4}} q_{2}^{4}.
\end{eqnarray}

{\em Rotating-wave approximation (RWA)}. We look for stationary modes of the form
$q_{1}(\tau)=q_{1}\ \sin(\Omega \tau),q_{2}(\tau)=q_{2}\ \sin(\Omega \tau)$, and use the approximation $\sin(x)^{3}\approx (3/4) \sin(x)$. We obtain the coupled equations
\begin{eqnarray}
-\Omega^{2} (q_{1}+\lambda q_{2}) + q_{1} -(3/4) \chi\ q_{1}^{3}&=&0\nonumber\\
-\Omega^{2} (q_{2}+\lambda q_{1}) + q_{2} -(3/4) \chi\ q_{2}^{3}&=&0\label{eq:4}
\end{eqnarray}
which is invariant under the change $q_{1}\rightarrow q_{2}$ and viceversa. Also if $(q_{1},q_{2})$ is a solution, so will $(-q_{1}, -q_{2})$. Let us examine some of the modes implied by Eqs.(\ref{eq:4}):\\
(i)\ $q_{1}=q_{2}\equiv q$ (Symmetric mode). This leads to $(1-\Omega^{2}(1+\lambda)) q - (3/4) \chi\ q^{3}=0$
which implies
\begin{equation}
q=0\hspace{1cm} or\hspace{1cm} q^{2}={(1-\Omega^{2}(1+\lambda))\over{(3/4)\chi}}
\end{equation}
and $\Omega^{2}<1/(1+\lambda)$ if $\chi>0$;otherwise $\Omega^{2}>1/(1+\lambda)$ if $\chi<0$.\\
(ii)\ $q_{2}=-q_{1}\equiv q$ (Antisymmetric mode). This leads to
$(1-\Omega^{2}(1-\lambda)) q - (3/4) \chi\ q^{3}=0$, which implies
\begin{equation}
q=0\hspace{1cm} or\hspace{1cm} q^{2}={(1-\Omega^{2}(1-\lambda))\over{(3/4)\chi}}
\end{equation}
and $\Omega^{2}<1/(1-\lambda)$ if $\chi>0$; otherwise $\Omega^{2}>1/(1-\lambda)$ if $\chi<0$.\\
(iii)\ $q_{1}^{2}\neq q_{2}^{2}$ (Asymmetric mode). After multiplying the first eq. in (\ref{eq:4}) by $q_{1}$ and the second eq. by $q_{2}$, and after subtracting, we obtain
\begin{equation}
(1-\Omega^{2}) = \left({3\over{4}}\right) \chi\ (q_{1}^{2}+q_{2}^{2})\label{eq:5}
\end{equation}
Next, we multiply the first eq. in (\ref{eq:4}) by $q_{2}$ and the second one by $q_{1}$, we obtain $(q_{1}^{2}-q_{2}^{2})[ \lambda \Omega^{2}- (3/4) \chi q_{1} q_{2} ]=0$. Then, assuming $q_{1}^{2}\neq q_{2}^{2}$, we obtain
\begin{equation}
\lambda \Omega^{2} = \left({3\over{4}}\right) \chi\ q_{1}\ q_{2}\label{eq:6}
\end{equation}
From Eqs.(\ref{eq:5}) and (\ref{eq:6}), we finally obtain
\begin{equation}
q_{1} = {1\over{2}} \left[\ \sqrt{{1-\Omega^{2}+2\lambda \Omega^{2}\over{(3/4) \chi}}} + \sqrt{{1-\Omega^{2}-2\lambda \Omega^{2}\over{(3/4) \chi}}}\ \right]\label{eq:q1}
\end{equation}
\begin{equation}
q_{2} = {1\over{2}} \left[\ \sqrt{{1-\Omega^{2}+2\lambda \Omega^{2}\over{(3/4) \chi}}} - \sqrt{{1-\Omega^{2}-2\lambda \Omega^{2}\over{(3/4) \chi}}}\ \right]\label{eq:q2}
\end{equation}
where, without loss of generality, we have assumed $q_{1}>q_{2}$.
\begin{figure}[t]
\noindent
\includegraphics[scale=0.75,angle=0]{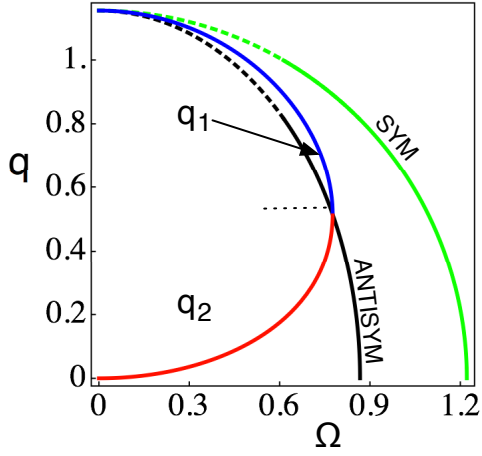}
\caption{Nonlinear mode amplitude as a function of  
mode frequency, for the symmetric, antisymmetric and asymmetric modes. Solid (dashed) portions denote stable (unstable) regimes. The solid curves labelled $q_{1}$, $q_{2}$ refer to the asymmetric mode. ($\chi=1, \lambda=0.33$)}
\label{fig}
\end{figure}
In order to have well-defined real solutions, we require $|\lambda|<1/2$ and
\begin{equation}
\Omega < \mbox{Min}\{{1\over{\sqrt{1-2\lambda}}}, {1\over{\sqrt{1+2\lambda}}}\} \hspace{1cm} \mbox{for}\hspace{1cm} \chi>0
\end{equation}
\begin{equation}
\Omega > \mbox{Max}\{{1\over{\sqrt{1-2\lambda}}}, {1\over{\sqrt{1+2\lambda}}}\} \hspace{1cm} \mbox{for}\hspace{1cm} \chi<0
\end{equation}

{\em Stability of the RWA modes}. Let us examine the linear stability of the RWA modes we just found. We set $q_{1}(\tau)= q_{1} \sin(\Omega \tau)+\delta q_{1}(\tau)$ and
$q_{2}(\tau)= q_{2} \sin(\Omega \tau)+\delta q_{2}(\tau)$, where $|\delta q_{1,2}|\ll q_{1,2}$. After inserting this into Eq.(\ref{eq:3}), keeping only linear terms in $\delta q_{1,2}$, and make use of the RWA for $\sin{x}^{2}\approx 1/2$,
we obtain the linear system:
\begin{equation}
{d^{2}\over{d \tau^{2}}} ( \delta q_{1} + \lambda \delta q_{2} ) + \left(1-{3\over{2}} \chi q_{1}^{2}\right) \delta q_{1} = 0
\end{equation}
\begin{equation}
{d^{2}\over{d \tau^{2}}} ( \delta q_{2} + \lambda \delta q_{1} ) + \left(1-{3\over{2}} \chi q_{2}^{2}\right) \delta q_{2} = 0
\end{equation}
We pose $\delta q_{1}(\tau)=\delta A \sin(\omega \tau)$ and $\delta q_{2}(\tau)=\delta B \sin(\omega \tau)$. The system has a nontrivial solution provided
\begin{equation}
(\omega^{2}-\alpha_{1})(\omega^{2}-\alpha_{2})-\lambda^{2}\omega^{4} = 0\label{eq:7}
\end{equation}
where $\alpha_{1,2}=1-(3/2) \chi q_{1,2}^{2}$.
From (\ref{eq:7}) we obtain
\begin{equation}
\omega^{2} = {\alpha_{1}+\alpha_{2} \pm \sqrt{(\alpha_{1}-\alpha_{2})^{2} + 4 \lambda^{2} \alpha_{1} \alpha_{2}}\over{2 (1-\lambda^{2})}}.\label{eq:8}
\end{equation}
Now, in order for the modes to be stable, we need $\omega^{2}>0$.\\
{\em Case $\chi<0$}.\  In this case, $\alpha_{1}=1+(2/3|\chi|)>0$ and $\alpha_{2}=1+(2/3|\chi|)>0$. The condition $\omega^2>0$, leads to $(\alpha_{1}-\alpha_{2})^2>-4 \lambda^2 \alpha_{1} \alpha_{2}$, which is always satisfied. Thus, for $\chi<0$ all modes are stable.\\
{\em Case $\chi>0$}. In this case, we need
\begin{equation}
q_{1}^{2} < {2\over{3 \chi}}, \ \ q_{2}^{2} < {2\over{3 \chi}} \hspace{0.5cm}\mbox{and} \hspace{0.5cm} |\lambda|<1.\label{eq:9}
\end{equation} 
\begin{figure}[t]
\noindent
\includegraphics[scale=0.45,angle=0]{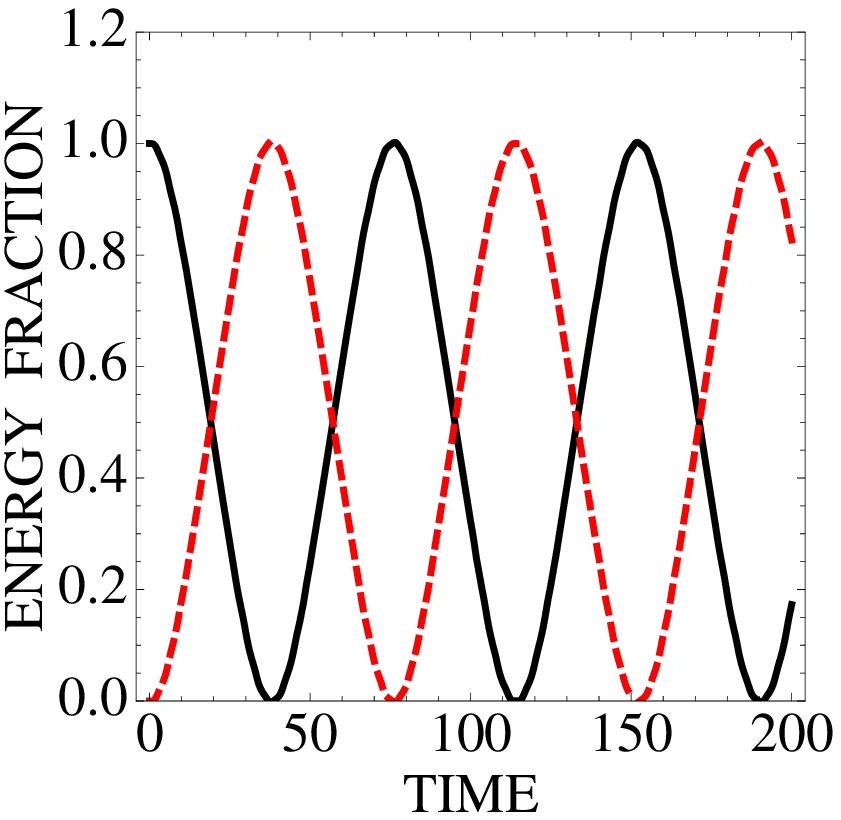}
\includegraphics[scale=0.45,angle=0]{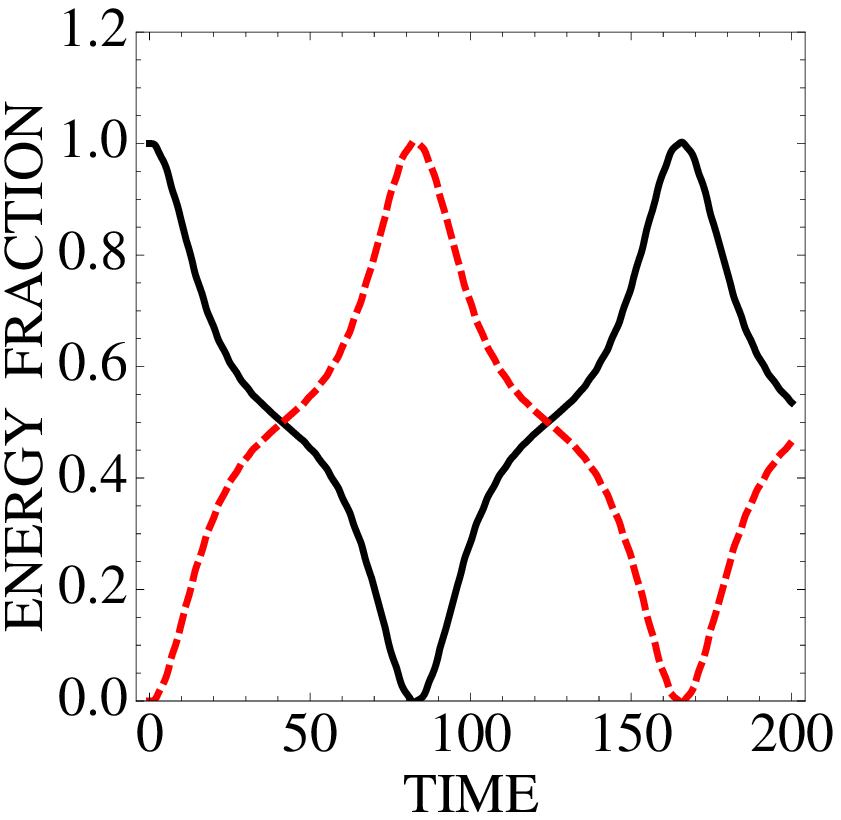}\\
\includegraphics[scale=0.45,angle=0]{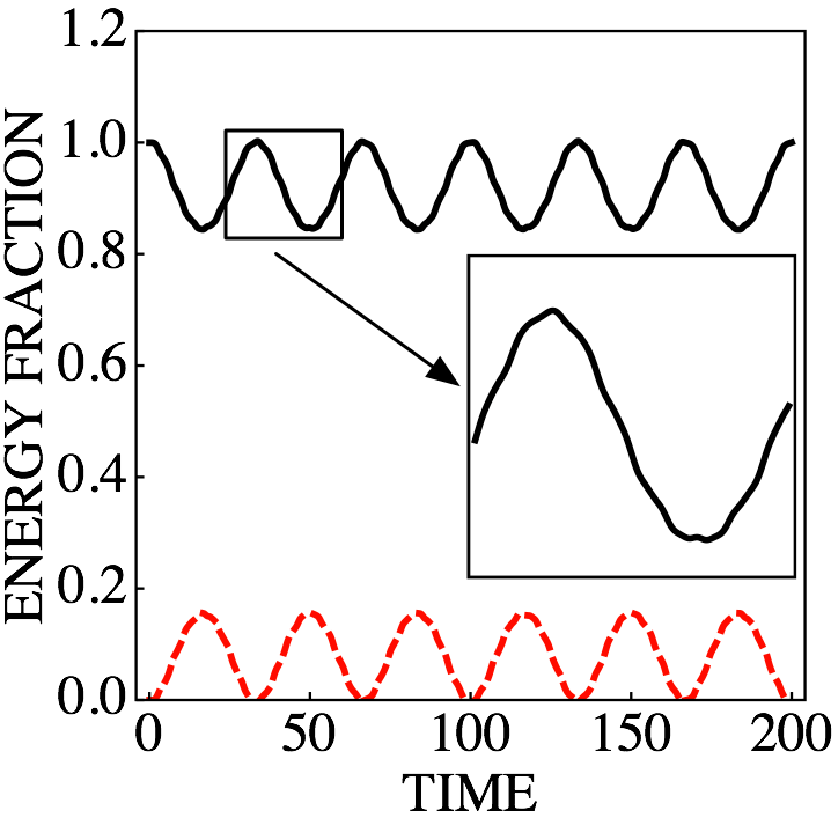}
\caption{Time evolution of individual SRR energies for increasing values of initial energies on one of the SRRs. $H_{1}(0)=0.73$ (left), $H_{1}(0)=1.0$ (right), $H_{1}(0)=1.2$ (bottom). The solid (dashed) curve denotes $H_{1} (H_{2})$. Inset shows small amplitude oscillations ($\chi=1/6, \lambda=0.1$)}
\label{fig3}
\end{figure}
Let us apply these conditions to the modes we found before.\\
(i) $q_{1}=q_{2}=\sqrt{ (1-\Omega^{2} (1+\lambda)/((3/4)\chi) }$. Condition (\ref{eq:9}) implies $\Omega^{2}>1/2 (1+\lambda)$, for stability. Combining this stability condition with the existence conditions, we conclude that the mode exists and is stable if
$1/(2 (1+\lambda)) < \Omega^{2} < 1/(1+\lambda)$.\\
(ii) $q_{2}=-q_{1}=\sqrt{ (1-\Omega^{2} (1-\lambda)/((3/4)\chi) }$. Condition
(\ref{eq:9}) implies $\Omega^{2}>1/2 (1-\lambda)$. Combining this stability condition with the existence conditions, we conclude that the mode exists and is stable if
$1/(2(1-\lambda)) < \Omega^{2} < 1/(1-\lambda)$.\\
(iii) $q_{1}^{2}\neq q_{2}^{2}$. Using (\ref{eq:q1}), (\ref{eq:q2}) and (\ref{eq:9}), one obtains the condition $\Omega^{2}>0$. Therefore, this mode is {\em always stable} in its existence domain, at least in the RWA framework.

{\em Dynamics}.\ An interesting question concerning the dynamics of the dimer is the selftraping problem: If we start with an initial condition where all the energy is residing on a single SRR, what are the conditions needed for that energy to remain ``localized'' or ``selftrapped'' on the initial SRR. An initial condition of the type $q_{1}(0)=q_{0}$, while $q_{2}(0)={\dot q_{1}(0)}={\dot q_{2}(0)}=0$, ensures $H_{1}(0)=H$, $H_{2}(0)=0$. Figure \ref{fig3} shows the time evolution (from Eq.(\ref{eq:3})) of the SRRs energies for three different values of initial energies, revealing a clear selftrapping transition, for initial energies above some threshold. Now, even though the curves resemble the ones observed for the cubic DNLS dimer, a close examination reveals the presence of small amplitude oscillations (not readily apparent in Fig.\ref{fig3}) around the main tendency. The initial-value problem seems, in fact, more difficult than its DNLS counterpart, and the possibility of achieving a closed-form solution seems doubtful at this stage, unless we find a way to average over the extra frequencies.

\begin{figure}[t]
\noindent
\includegraphics[scale=0.5,angle=0]{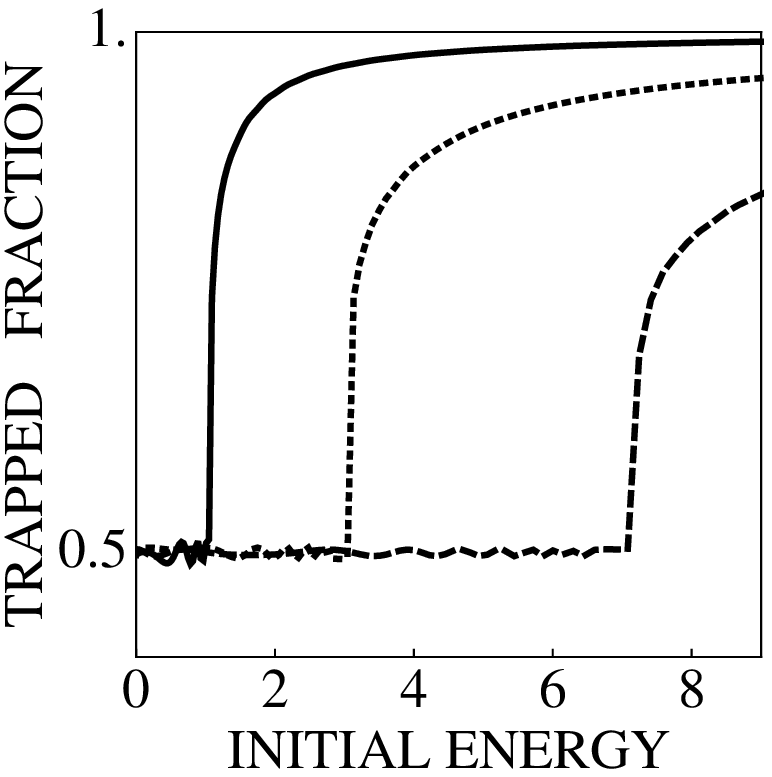}\hspace{0.2cm}
\includegraphics[scale=0.45,angle=0]{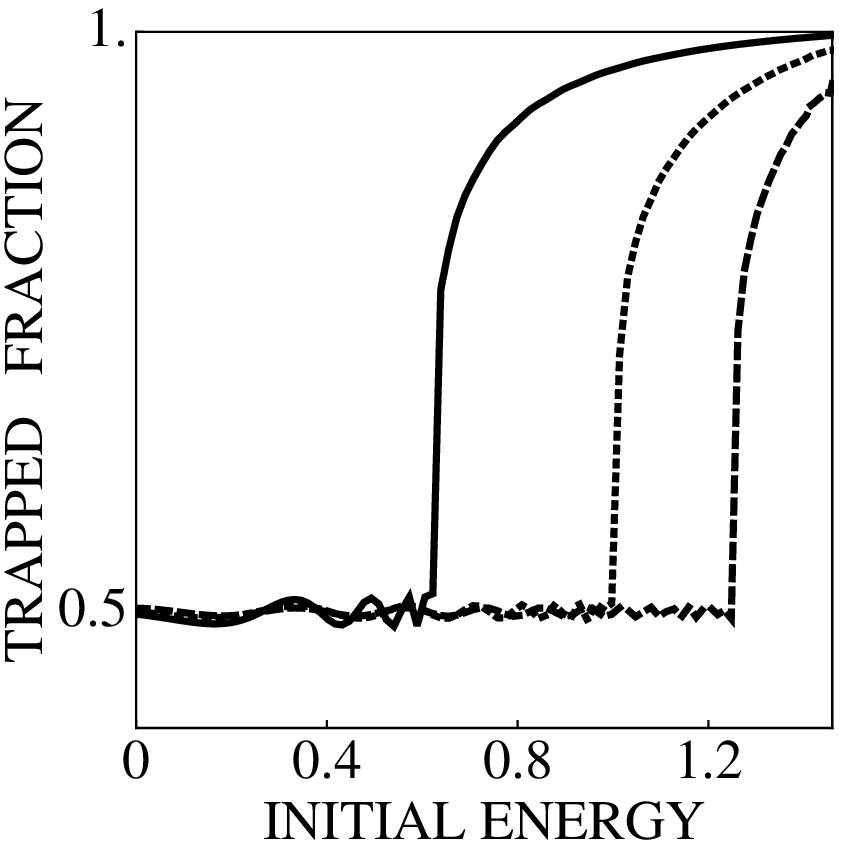}
\caption{Long-time averaged fraction of initial energy residing on SRR as a function of its initial energy, for several coupling parameter values
: $\lambda=0.05$(solid), $\lambda=0.1$ (dotted) and $\lambda=0.15$ (dashed). Left: Hard nonlinearity ($\chi=-1/6$). Right: Soft nonlinearity ($\chi=1/6$).}
\label{fig4}
\end{figure}

\begin{figure}[h]
\noindent
\includegraphics[scale=0.45,angle=0]{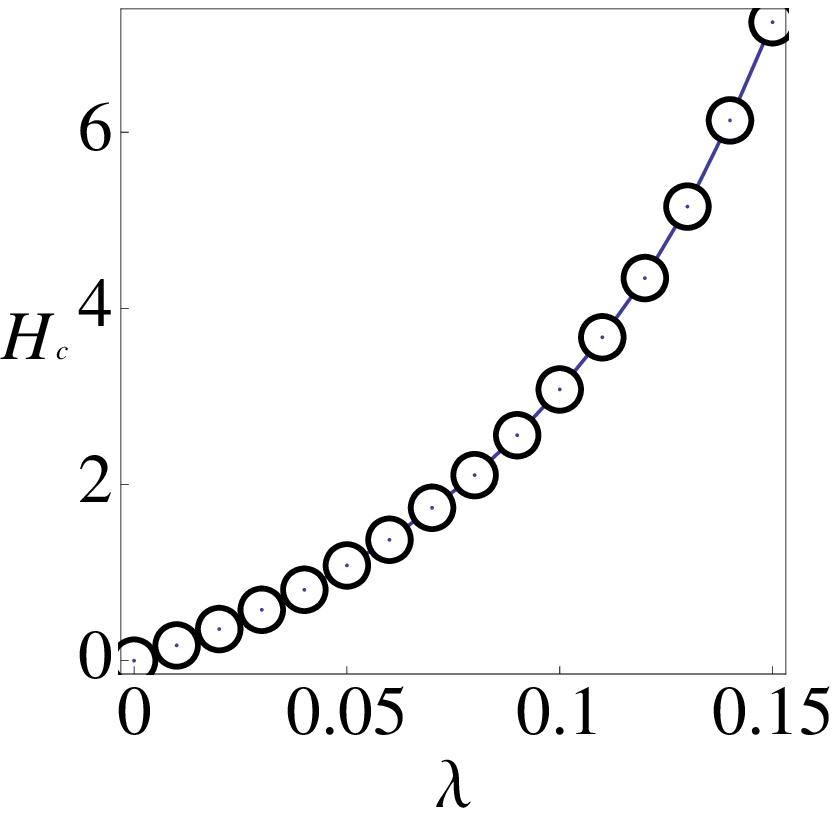}\hspace{0.4cm}
\includegraphics[scale=0.45,angle=0]{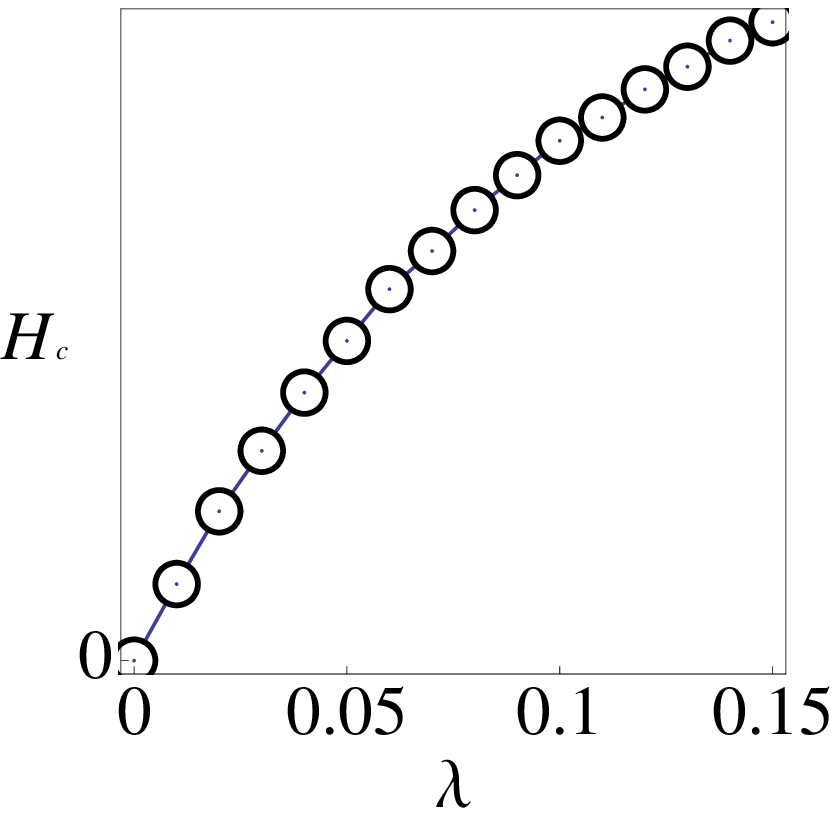}
\caption{Minimum initial energy on a SRR to begin selftraping, as a function of the value of inductive coupling. Left: Hard nonlinearity ($\chi=-1/6$). Right: Soft nonlinearity ($\chi=1/6$). }
\label{fig5}
\end{figure}

Figure \ref{fig4} shows the long-time averaged fraction of energy remaining on the initial SRR, as a function of its initial energy, for hard and soft nonlinearities, and for different mutual inductance couplings.
Clearly, for a given value of the nonlinearity parameter, the threshold energy needed for selftrapping is an increasing function of the inductive coupling $\lambda$. This information is conveyed in a more clear manner in Fig.\ref{fig5} where we show the threshold value of $H_{1}(0)=H$ needed to effect selftrapping as a function of the inductive coupling parameter. For both cases, the ``soft'' and the ``hard'' nonlinearity cases, the minimum initial energy is a increasing function of coupling, as expected on general grounds, but with different curvatures: 
Positive for hard nonlinearity, while negative 
for soft nonlinearity.

{\em Conclusions}. We have computed in closed form the stationary modes of a nonlinear magnetoinductive dimer in the rotating-wave approximation. The linear stability window for the symmetric, antisymmetric and asymmetric modes was obtained  in closed form, finding that the asymmetric mode 
is always stable. The dynamics evolution of a localized initial excitation reveals a selftrapping transition, with an energy threshold that increases with an increase in magnetic coupling.

This work was supported in part by Fondo Nacional de Ciencia y Tecnolog\'{\i}a (Grant 1120123), Programa Iniciativa Cient\'{\i}fica Milenio (Grant P10-030-F), and Programa de Financiamiento Basal (Grant FB0824/2008).

\end{document}